\documentclass[graybox]{svmult}

\usepackage{mathptmx}       % selects Times Roman as basic font   maha se za po-krasivo ...
\usepackage{helvet}         % selects Helvetica as sans-serif font
\usepackage{courier}        % selects Courier as typewriter font
\usepackage{type1cm}        % activate if the above 3 fonts are
\usepackage{makeidx}         % allows index generation
\usepackage{graphicx}        % standard LaTeX graphics tool
\usepackage{multicol}        % used for the two-column index
\usepackage[bottom]{footmisc}% places footnotes at page bottom

\newcommand\be{\begin{equation}}
\newcommand\ee{\end{equation}}
\newcommand{\bea}{\begin{eqnarray}}
\newcommand{\eea}{\end{eqnarray}}

\newcommand{\nn}{\nonumber}
\newcommand{\pd}{\partial}

\begin{document}

\title*{Glueball Inflation and Gauge/Gravity Duality}
% Use \titlerunning{Short Title} for an abbreviated version of
% your contribution title if the original one is too long
\author{Lilia Anguelova}
% Use \authorrunning{Short Title} for an abbreviated version of
% your contribution title if the original one is too long
\institute{Lilia Anguelova \at Institute for Nuclear Research and Nuclear Energy, Bulgarian Academy of Sciences, Sofia 1784, Bulgaria, \email{anguelova@inrne.bas.bg}}
%\and Name of Second Author \at Name, Address of Institute \email{name@email.address}}
%
%
\maketitle

\abstract{We summarize our work on building glueball inflation models with the methods of the gauge/gravity duality. We review the relevant five-dimensional consistent truncation of type IIB supergravity. We consider solutions of this effective theory, whose metric has the form of a $dS_4$ foliation over a radial direction. By turning on small (in an appropriate sense) time-dependent deformations around these solutions, one can build models of glueball inflation. We discuss a particular deformed solution, describing an ultra-slow roll inflationary regime.}

\section{Introduction}
\label{sec:1}

Composite inflation models \cite{CJS,BCJS} provide a possible resolution to the well-known $\eta$-problem \cite{CLLSW,DRT} of inflationary model-building. However, they are quite challenging to study with standard QFT methods, since they involve a strongly-coupled gauge sector. This has motivated interest in developing descriptions of such models via a string-theoretic tool aimed precisely at studying the nonperturbative regime of gauge theories, namely the gauge/gravity duality. Gravitational duals, in which the inflaton arises from the position of a D3-brane probe have been considered in \cite{AB1,AB2,AB3,AB4,EFK}. Instead, in \cite{ASW,ASW2,LA} we studied models, whose inflaton arises from the background fields of the gravitational solution and is thus a glueball in the dual gauge theory.

The backgrounds of interest for us solve the equations of motion of the 5d consistent truncation of type IIB supergravity established in \cite{BHM}. The latter encompasses a wide variety of prominent gravity duals, like \cite{MN,KS,NPP,ENP,EGNP}, as special solutions and thus provides a unifying framework for gauge/gravity duality investigations. The work \cite{ASW} obtained new non-supersymmetric classes of solutions of this theory, whose metric is of the form of a $dS_4$ fibration over the fifth direction. These backgrounds provide a useful playground for studying certain strongly-coupled gauge theories in de Sitter space. To have an inflationary model, however, one needs a time-dependent Hubble parameter. Therefore, in \cite{LA} we investigated time-dependent deformations around a solution of \cite{ASW}, in order to search for gravity duals of glueball inflation.   

It is worth pointing out that the main cosmological observables of an inflationary model (like the scalar spectral index $n_s$ and the tensor-to-scalar ratio $r$) are entirely determined by the Hubble parameter and inflaton field as functions of time \cite{DB}. Hence, once one has a deformed background in the above set-up, one can immediatey compute the desired quantities. This is the sense, in which the time-dependent deformations of the previous paragraph give models of cosmological inflation. In that vein, in \cite{LA} we calculated the slow roll parameters for a solution we found there and thus established that it gives a gravity dual of ultra-slow roll glueball inflation. The ultra-slow roll regime may play an important role in understanding the observed low $l$ anomaly in the power spectrum of the CMB. Hence it deserves further study. We also discuss here perspectives for building gravity duals of standard slow roll inflationary models.

\section{Effective 5d theory}

In this section we summarize necessary material about the 5d consistent truncation of type IIB supergravity relevant for our considerations. We also recall a particular solution of this theory, whose time-dependent deformations we will investigate in the next subsection.

\subsection{Action and field equations}

Let us briefly review the basic characteristics of the five-dimensional consistent truncation of \cite{BHM}. Using a particular ansatz for the bosonic fields of type IIB supergravity in terms of certain 5d fields and integrating out five compact dimensions, one reduces the ten-dimensional IIB action to the following five-dimensional one:
\be \label{5dAction}
S = \int d^5 x \sqrt{- det g} \left[ - \frac{R}{4} + \frac{1}{2} G_{ij} (\Phi) \pd_I \Phi^i \pd^I \Phi^j + V (\Phi) \right] \,\, .
\ee
Here $\{ \Phi^i \}$ is a set of 5d scalar fields, that arise from the components of the 10d ones including metric warp factors, $V(\Phi)$ is a rather complicated potential, $G_{ij} (\Phi)$ is a diagonal sigma-model metric and, finally, $R$ is the Ricci scalar of the 5d spacetime metric $g_{IJ}$. The full expressions for $V(\Phi)$ and $G_{ij} (\Phi)$ can be found in \cite{BHM}; for a more concise summary, see also \cite{ASW}. The field equations that the action (\ref{5dAction}) implies are:
\bea \label{EoM}
\nabla^2 \Phi^i + {\cal G}^i{}_{jk} \,g^{I J} (\pd_I \Phi^j) (\pd_J \Phi^k) -V^i &=& 0 \quad , \nn \\
- R_{IJ} + 2 \,G_{ij} \,(\pd_I \Phi^i) (\pd_J \Phi^j) + \frac{4}{3} \,g_{IJ} V &=& 0 \quad ,
\eea
where $V^i = G^{ij} V_j$ \,, \,$V_i = \frac{\pd V}{\pd \Phi^i}$ \,and ${\cal G}^i{}_{jk}$ are the Christoffel symbols of the sigma-model metric $G_{ij}$\,.

\subsection{A solution with $dS_4$ slicing}

We will be interested in time-dependent deformations around a particular solution of the system (\ref{EoM}) found in \cite{ASW}. So let us first recall its form. In the notation of \cite{ASW} and working within the same subtruncation as there (i.e., with zero NS flux), we have six scalars in the 5d effective theory:
\be 
\{\Phi^i (x^I)\} = \{ p (x^I), x (x^I), g (x^I), \phi (x^I), a(x^I), b(x^I) \} \,\, .
\ee
The work \cite{ASW} found three families of solutions of (\ref{EoM}) with a 5d metric of the form
\be \label{metricans}
ds_5^2 = e^{2A(z)} \left[ -dt^2 + s(t)^2 \sum_{m=1}^{3} (dx^m)^2 \right] + dz^2 \,\, ,
\ee
where $s(t) = e^{{\cal H}t}$ with ${\cal H} = const$. In all of them, three of the scalars $\Phi^i$ vanish identically, namely:
\be
g (x^I) = 0 \quad , \quad a (x^I) = 0 \quad , \quad b (x^I) = 0 \quad .
\ee
Two of those solutions are numerical and one is analytical. For convenience, we will study deformations around the latter. Denoting its metric functions and scalar fields by the subscript $0$, we have \cite{ASW}:
\bea \label{ZeroOrSol}
A_0 (z) &=& \ln (z+C) + \frac{1}{2} \,\ln \!\left( \frac{7}{3} {\cal H}_0^2 \right) \,\, , \nn \\
p_0 (z) &=& - \frac{1}{7} \ln (z+C) - \frac{1}{14} \,\ln \!\left( \frac{7 N^2}{9} \right) \,\, , \nn \\
x_0 (z) &=& - 6 \,p_0 (z) \qquad , \qquad \phi_0 = 0 \quad ,
\eea
where $C$ and $N$ are constants.

Let us mention in passing that the form of (\ref{ZeroOrSol}) is consistent with ALD (asymptotically linear dilaton) behavior at large $z$. This is not obvious at first sight due to the use of a different coordinate system (in string frame) compared to the conventional one (in Einstein frame), in which the holographic renormalization of ALD backgrounds was developed \cite{MMcN}. This issue was discussed in more detail in \cite{ASW,ASW2}, where it was also pointed out that the same kind of asymptotics characterizes the walking solutions of \cite{NPP} as well.

\section{Deforming the $dS_4$ solution}

Now we are ready to turn to the investigation of solutions of the system (\ref{EoM}), which are deformations around the zeroth order background (\ref{ZeroOrSol}). Since our aim is to study glueball inflation, we would like to find solutions, whose 5d metric is of the form (\ref{metricans}) but with $\dot{\cal H} \neq 0$. Recall that the Hubble parameter is defined as
\be \label{Hdef}
{\cal H} = \frac{\dot{s}}{s} \quad ,
\ee
where for convenience we have denoted \,$\dot{} \equiv \frac{\pd}{\pd t}$\,. \,Now, one of the slow roll conditions widely used in inflationary model building\footnote{One should keep in mind, though, that there are more exotic inflationary regimes, in which one or more of the slow roll conditions can be violated; see \cite{AL,MMS,MSY}, for instance.} is the following \cite{DB}:
\be \label{SRcond}
- \frac{\dot{{\cal H}}}{{\cal H}^2} \,<\!\!< \,1 \,\,\,.
\ee
In view of that, we will look for solutions with time-dependent ${\cal H}$ by considering small, in the sense of (\ref{SRcond}), deformations around an ${\cal H} = const$ solution.

For that purpose, let us introduce a small parameter $\gamma$, satisfying
\be
\gamma <\!\!< 1 \,\, ,
\ee
and search for solutions that are expansions in powers of this parameter. To do this, we make the following ansatz for the nonvanishing 5d fields:
\bea \label{Expansions}
p (t,z) &=& p_0 (z) \quad , \quad x (t,z) = x_0 (z) \,\,\, , \nn \\
\phi(t,z) &=& \gamma \,\phi_{(1)} (t,z) + \gamma^3 \phi_{(3)} (t,z) + {\cal O} (\gamma^5) \,\,\, , \nn \\
A(t,z) &=& A_0(z) + \gamma^2 A_{(2)}(t,z) + {\cal O} (\gamma^4) \,\,\, , \nn \\
H(t,z) &=& {\cal H}_0 \,t + \gamma^2 H_{(2)} (t,z) +{\cal O} (\gamma^4) \,\,\, ,
\eea
where $H(t,z)$ is a warp factor defined via
\be \label{TimeDepMetricAnz}
ds_5^2 = e^{2A(t,z)} \left[ -dt^2 + e^{2H(t,z)} \sum_{m=1}^{3} (dx^m)^2 \right] + dz^2 \,\,\, .
\ee
In other words, we keep the scalars $p (x^I)$ and $x (x^I)$ the same as in (\ref{ZeroOrSol}), while allowing small deviations around that zeroth order solution in the scalar $\phi$ and the metric functions $A$ and $H$.

It is worth commenting a bit more on the form of the deformation ansatz (\ref{Expansions}). First of all, in order to obtain solutions with $\dot{{\cal H}} \neq 0$, we need to turn on time dependence in at least one scalar. It is convenient to take this scalar to be $\phi$ since, unlike $p$ and $x$, it vanishes at zeroth order and, furthermore, it is a flat direction of the potential; see \cite{LA}. Therefore, $\phi$ will play the role of the inflaton in our set-up. Also note that, although we would like to have $t$-dependent $H$ only, we have allowed $z$-dependence too, for more generality. And, finally, the different powers of $\gamma$ in the expansion of $\phi$, compared to the expansions of the warp factors, will be of great significance for finding an analytical solution, as will become clear below.

\subsection{Equations of motion}
\label{subsec:3.1}

Let us now substitute the ansatz (\ref{Expansions}) in the system (\ref{EoM}) and study the result order by order in $\gamma$. Clearly, since we are expanding around a zeroth order solution, there is no contribution at order $\gamma^0$.

At order $\gamma$, we have the following field equation \cite{LA}:
\be \label{phi1}
\ddot{\phi}_{(1)} + 3 \,{\cal H}_0 \,\dot{\phi}_{(1)} = e^{2 A_0} \!\left( \phi''_{(1)} + 4 A'_0 \phi'_{(1)} \right) \,\, ,
\ee
where $'\equiv \frac{\pd}{\pd z}$\,. To find a solution, let us make the ansatz
\be
\phi_{(1)} = \Phi_1 (t) \,\Phi_2 (z)
\ee
and solve the eigen problems
\be \label{EigenTZ}
\ddot{\Phi}_1 + 3 {\cal H}_0 \dot{\Phi}_1 = \lambda \,\Phi_1 \qquad {\rm and} \qquad e^{2 A_0} \!\left( \,\Phi''_2 + 4 A'_0 \Phi'_2 \,\right) = \lambda \,\Phi_2 
\ee
with $\lambda$ being some constant. One easily obtains that \cite{LA}:
\be \label{Phi1tsol}
\Phi_1 (t) = C_1 \,e^{k_+ t} + C_2 \,e^{k_- t} \,\, , \,\,\, {\rm where} \,\,\,\, k_{\pm} = -\frac{3 {\cal H}_0}{2} \pm \frac{\sqrt{9 {\cal H}_0^2 + 4 \lambda} }{2}
\ee
and $C_{1,2}$ are integration constants, while
\be \label{PhizC3C4Sol}
\Phi_2 (z) = C_3 (z+C)^{\alpha_+} + C_4 (z+C)^{\alpha_-} \,\,\, {\rm with} \,\,\,\, \alpha_{\pm} = - \frac{3}{2} \pm \frac{3}{2} \sqrt{1 + \frac{4 \,\lambda}{21 \,{\cal H}_0^2}}
\ee
and $C_{3,4}$ being integration constants.

Note that if $\lambda = 0$, then one is free to add an arbitrary constant to the $\phi_{(1)}$ solution, determined by (\ref{EigenTZ}). This will be important in the following.

At order $\gamma^2$, we find a coupled system for the warp factor deformations $A_{(2)}$ and $H_{(2)}$, namely \cite{LA}:
\bea \label{E1E2E3E4}
&&E1: \quad - \,{\cal H}_0^2 \!\left( \frac{7}{3} (z+C)^2 A_{(2)}'' + \frac{56}{3} (z+C) A_{(2)}' + 7 (z+C) H_{(2)}' + 6 A_{(2)} \right) \nn \\
&&\hspace*{1.1cm}+ \,{\cal H}_0 \!\left( 3 \dot{A}_{(2)} + 6 \dot{H}_{(2)} \right) + 3 \ddot{A}_{(2)} + 3 \ddot{H}_{(2)} + \frac{1}{2} \dot{\phi}^2_{(1)} = 0 \,\,\,\, , \nn \\
&&E2: \quad {\cal H}_0^2 \!\left( \frac{7}{3} (z+C)^2 \left[ A_{(2)}'' + H_{(2)}'' \right] + \frac{56}{3} (z+C) A_{(2)}' + \frac{49}{3} (z+C) H_{(2)}' \right. \nn \\
&&\hspace*{1.1cm}+ \,6 A_{(2)} \bigg) - {\cal H}_0 \!\left( 5 \dot{A}_{(2)} + 6 \dot{H}_{(2)} \right) - \ddot{A}_{(2)} - \ddot{H}_{(2)} = 0 \,\,\,\, , \nn \\ 
&&E3: \quad 4 A_{(2)}'' + 3 H_{(2)}'' + \frac{2}{z+C} \left( 4 A_{(2)}' + 3 H_{(2)}' \right) + \frac{1}{2} \phi'^{\,2}_{(1)} = 0 \,\,\,\, , \nn \\
&&E4: \quad 3 \dot{A}_{(2)}' + 3 \dot{H}_{(2)}' + 3 \,{\cal H}_0 \,H_{(2)}' + \frac{1}{2} \dot{\phi}_{(1)} \phi'_{(1)} = 0 \,\,\,\, .
\eea
To solve this rather involved system, let us take for convenience the $\phi_{(1)}$ solution to be:
\be \label{phi1withaddconst}
\phi_{(1)} = C_{\phi} + \tilde{C} e^{kt} (z+C)^{\alpha} \qquad {\rm with} \qquad C_{\phi}\,, \,\tilde{C} = const \,\,\,\, ,
\ee
where $k$ is any of $k_{\pm}$ and $\alpha$ is any of $\alpha_{\pm}$\,. Note that the addition of the arbitrary constant $C_{\phi}$ in (\ref{phi1withaddconst}) makes no difference for the solutions of (\ref{E1E2E3E4}), since the function $\phi_{(1)}$ enters those equations only through its derivatives. However, the presence of $C_{\phi}$ will turn out to be useful later. Plus, it will become clear shortly that it is consistent with (\ref{EigenTZ}).

Now, the form of $E3$ in (\ref{E1E2E3E4}), together with (\ref{phi1withaddconst}), suggests looking for a solution with the following ansatz:
\be \label{AHans}
A_{(2)} (t,z) = e^{2 k t} \hat{A}(z) \qquad {\rm and} \qquad H_{(2)} (t,z) = \hat{C}_H + e^{2kt} \hat{H}(z) \,\, ,
\ee
where $\hat{C}_H = const$. Again, we have included an arbitrary constant $\hat{C}_H$, since $H_{(2)}$ enters the system (\ref{E1E2E3E4}) only via its derivatives. Substituting (\ref{AHans}) and (\ref{phi1withaddconst}) into (\ref{E1E2E3E4}), one can see that the time-dependence factors out. Thus, one is left with a coupled system of ODEs for the functions $\hat{A} (z)$ and $\hat{H} (z)$. A detailed investigation in \cite{LA} showed that this system has a solution only for\footnote{To prove this, one also needs to use the fact that (\ref{Phi1tsol}) and (\ref{PhizC3C4Sol}) imply the following relation between $k$ and $\alpha$: \, $k = - \frac{3}{2} {\cal H}_0 \pm \frac{\sqrt{84 \alpha^2 + 252 \alpha + 81}}{6} \,{\cal H}_0$ \,.}
\be
\alpha = 0 \,\,\, ,
\ee
in which case both $\hat{A} = const$ and $\hat{H} = const$. Substituting $\alpha = 0$ in (\ref{PhizC3C4Sol}), we find that $\lambda = 0$ as well. This in turn implies that we are free to add the constant $C_{\phi}$ in (\ref{phi1withaddconst}), as commented below (\ref{PhizC3C4Sol}). Finally, from (\ref{Phi1tsol}) we now have:
\be
k = - 3 {\cal H}_0 \,\,\, ,
\ee
where we have taken the value of $k_-$ in order to have time-dependence in the inflaton field $\phi$.

\subsection{Ultra-slow roll inflation}

The solution we described above gives a dual description of an ultra-slow roll glueball inflation model. To see this, let us compute the inflationary slow-roll parameters. They are defined in terms of the inflaton field and Hubble parameter as \cite{DB}:
\be \label{epetaH}
\varepsilon = - \frac{\dot{{\cal H}}}{{\cal H}^2} \qquad {\rm and} \qquad \eta = - \frac{\ddot{\phi}}{{\cal H} \dot{\phi}} \qquad .
\ee
From the results of Subsection \ref{subsec:3.1}, we have that $\phi$ and ${\cal H}$ are given by:
\bea \label{Hphi}
\phi &=& \left( C_{\phi} + \tilde{C} \,e^{-3 {\cal H}_0 t} \right) \gamma + {\cal O} (\gamma^3) \,\, , \nn \\
{\cal H} &=& {\cal H}_0 - C_{\cal H} \,e^{-6 {\cal H}_0 t} \,\gamma^2 + {\cal O} (\gamma^4) \,\, ,
\eea
where $C_{\cal H}$ is some constant; for more details, see \cite{LA}.\footnote{Note that, since the correction $A_{(2)}$ to the warp factor $A(t,z)$ in (\ref{TimeDepMetricAnz}) also depends on $t$, as can be seen from (\ref{AHans}), one should, in principle, first perform a coordinate transformation $t \rightarrow \tau$ that absorbs that dependence, before computing the physical Hubble parameter ${\cal H} (\tau)$ and inflaton field $\phi (\tau)$. However, in the present case, this leads to exactly the same expressions as (\ref{Hphi}) with $t$ substituted by $\tau$, with the only difference being the numerical value of the constant $C_{\cal H}$. So we will not discuss the details of that transformation here.}

Substituting (\ref{Hphi}) in (\ref{epetaH}), we find that the slow roll parameters behave as $\varepsilon = {\cal O} (\gamma^2)$ and $\eta = 3 + {\cal O} (\gamma^2)$; see \cite{LA} for more detailed expressions. In other words, at leading order we have:
\be
\varepsilon <\!\!< 1 \qquad {\rm and} \qquad \eta = 3 \quad .
\ee
These are precisely the values of $\varepsilon$ and $\eta$ for the ultra-slow regime, considered in \cite{TW,WK}. In fact, our result for the inflaton in (\ref{Hphi}) also agrees completely with the expression in \cite{WK}.

It is worth pointing out a similarity between our model and the constant-rate-of-roll solutions of \cite{MSY}. For that purpose, let us introduce the following series of slow roll parameters:
\be
\varepsilon_1 = - \frac{\dot{{\cal H}}}{{\cal H}^2} \qquad {\rm and} \qquad \varepsilon_{n+1} = \frac{\dot{\varepsilon}_n}{{\cal H}\varepsilon_n} \quad ,
\ee
where obviously $\varepsilon_1 \equiv \varepsilon$. One can easily compute that, at large $t$, our solution gives \cite{LA}:
\be
\varepsilon_{2n+1} \rightarrow 0 \qquad {\rm and} \qquad \varepsilon_{2n} \rightarrow - 6 \quad .
\ee
This is consistent with the asymptotics in \cite{MSY}. It would be interesting to investigate whether there is a deeper underlying reason for that.

In conclusion, let us make a few comments regarding other inflationary models in our framework. Although an ultra-slow roll inflationary regime may be desirable to account for the low $l$ anomaly in the CMB power spectrum, it is rather short-lived. So it has to be succeeded by regular slow roll, in order to have enough expansion and thus give a complete inflationary model. To obtain such solutions in our gauge/gravity duality set-up, one may need to study deformations around the numerical solutions of \cite{ASW}, instead of the analytical one (\ref{ZeroOrSol}). It could also be that duals of regular slow roll can be found by modifying the initial ansatz for the deformations around the analytical solution. Finally, it would be interesting to investigate what kind of models can be obtained by going to the next order in $\gamma$ in the expansions (\ref{Expansions}), while taking $\phi_{(1)}$, $A_{(2)}$ and $H_{(2)}$ to vanish. This seems to open much wider possibilities for inflationary model building, as the equations of motion for $A_{(4)}$ and $H_{(4)}$ would be independent of $\phi_{(3)}$. Thus, many of the restrictions we encountered here (and as a result of which we ended up with ultra-slow roll) would not occur.

\bigskip

\begin{acknowledgement}
I have received partial support from the European COST Action MP-1210 and the Bulgarian NSF grant DFNI T02/6.
\end{acknowledgement}
\end{document}